# China daily $CO_2$ emission estimates 2019–2025

Xiaoting Huang[1,2], Zhu Deng[3,4], Ling Ye[5], Peng Lu[6], Xuanren Song[1,2], Yuli Shan[7,8,9], Zhu Liu[1,2]

[1] State Key Laboratory of Hydroscience and Engineering, Tsinghua University, Beijing 100084, China

[2] Department of Earth System Science, Ministry of Education Key Laboratory for Earth System Modeling, Institute for Global Change Studies, Tsinghua University, Beijing 100084, China

[3] Department of Geography, The University of Hong Kong, Hong Kong SAR, China

[4] Institute for Climate and Carbon Neutrality, The University of Hong Kong, Hong Kong SAR, China

[5] College of Geo-Exploration Science and Technology, Jilin University, Changchun 130026, China

[6] Dayu College, Hohai University, Nanjing 210098, China

[7] State Key Laboratory of Water Pollution Control and Green Resource Recycling, School of the Environment, Nanjing University, Nanjing 210023, China

[8] Institute for the Environment and Health, Nanjing University Suzhou Campus, Suzhou 215163, China

[9] School of Geography, Earth and Environmental Sciences, University of Birmingham, Birmingham B15 2TT, UK

**Corresponding author(s):** Zhu Liu**, E-mail:** zhuliu@tsinghua.edu.cn

## Abstract

Near-real-time $CO_2$ estimates enable timely tracking of emission changes, but broad sectoral classifications limit their ability to resolve industrial dynamics across China's provinces. Here, we update Carbon Monitor to estimate daily $CO_2$ emissions across five sectors in 31 mainland Chinese provinces during 2019–2025 and monthly emissions from 40 industrial and construction subsectors in 30 provinces, combining historical inventories with industrial production indicators. National emissions increased from 10.83 Gt $CO_2$ in 2019 to 12.15 Gt $CO_2$ in 2023, before declining slightly to 12.08 Gt $CO_2$ in 2025. The 2024–2025 national reduction of 0.23% reflected opposing sectoral trends, with industrial emissions increasing by 48.8 Mt $CO_2$ while power generation and ground transport emissions decreased by 46.6 Mt $CO_2$ and 27.7 Mt $CO_2$, respectively. Provincial trajectories also diverged, with emissions rising in 16 provinces and falling in 15. Industrial disaggregation revealed increases of 102.3 Mt $CO_2$ in nonferrous-metal smelting and pressing and decreases of 108.1 Mt $CO_2$ in nonmetal mineral products during 2019–2024. Comparison with a historical provincial inventory yielded a median absolute relative difference of 9.4%. These findings demonstrate that national emission stabilization masks divergent regional and industrial changes, highlighting the need for temporally updated, sector-specific evidence to assess progress toward sustained reductions and identify mitigation priorities.

**Keywords: provincial $CO_2$ emissions, Carbon Monitor, China, industrial subsectors, emission inventories**

## 1 Introduction

Rapid and sustained reductions in anthropogenic carbon dioxide ($CO_2$) emissions are required to limit global warming in accordance with the Paris Agreement [1,2]. Continued fossil-fuel emissions reduce the remaining carbon budget, making the timing and persistence of emission reductions central to climate mitigation [3–5]. China's commitments to peak $CO_2$ emissions before 2030 and achieve carbon neutrality before 2060 increase the importance of tracking its evolving emission trajectory. Assessing progress requires information on emission levels, the persistence of recent changes, and the sectors and regions contributing to those changes [6–8]. Conventional inventories provide detailed accounting based on energy statistics and emission factors [9,10], but their compilation requires the collection and reconciliation of activity data, resulting in delays in the availability of emission estimates for recent periods. Uncertainties in energy statistics and emission factors can also lead to substantial discrepancies in estimated $CO_2$ emissions [11–13]. Frequently updated estimates complement this statistical foundation by identifying emerging changes before complete inventories become available [14].

Timely emission information is also needed at the provincial level. National totals combine different regional trajectories and can conceal substantial increases and decreases across provinces and sectors that offset in aggregation [15]. China's provinces differ in resource endowments, industrial specialization, electricity supply, transport activity, and heating demand [16–18]. Previous studies have documented spatial disparities in provincial emission intensity, regional emission patterns, and per-capita emissions [19–22], while sectoral inventories, emission decomposition analyses, and energy-system modeling have examined the roles of electricity generation, industrial activity, and regional energy structures [18,23–26]. Large emitting provinces can strongly influence the national balance, whereas rapid percentage growth in smaller provinces may indicate emerging shifts despite modest absolute contributions [18,20]. Examining provincial trajectories together with sectoral contributions is therefore necessary to distinguish broadly shared trends from opposing changes concealed by a small national net change, and to identify where these changes occur.

Existing inventories provide the foundation for this assessment. China Emission Accounts and Datasets (CEADs) offer annual provincial $CO_2$ accounts with detailed fuel and source classifications [9,27,28]. These accounts provide emission estimates for 47 socioeconomic sectors, enabling detailed assessments of emissions across different economic activities. Complementary inventories, including MEIC-global-$CO_2$, the Emissions Database for Global Atmospheric Research (EDGAR), the Open-source Data Inventory for Anthropogenic $CO_2$ (ODIAC), and the China High Resolution Emission Database (CHRED), provide source-specific and spatially detailed emission estimates using different statistical inputs, point-source information, and spatial allocation approaches [29–34]. More recent datasets have extended emission accounting to monthly county-level estimates and city inventories with broader accounting boundaries [35,36]. However, comparisons among inventories have revealed substantial discrepancies associated with differences in activity data, emission factors, accounting boundaries, and spatial allocation methods [37–40]. Although these inventories provide valuable information for historical and spatial analyses, many depend on compiled activity statistics that constrain the timely assessment of recent emission changes [10]. There is therefore a need for emission estimates that combine timely updates with sufficiently detailed sectoral information to support more differentiated assessments of emission dynamics.

Near-real-time frameworks address the temporal gap by combining emission baselines with rapidly updated activity indicators. Carbon Monitor established daily emission estimates across major emitting sectors [41,42], while related studies have extended high-frequency monitoring to Chinese provinces, cities worldwide, and European countries [43–46]. Satellite-based approaches have also been explored to constrain daily anthropogenic $CO_2$ emissions using atmospheric observations of air pollutants [47]. Daily and provincial analyses of the COVID-19 period showed how short-term changes in mobility and production could be detected before complete energy statistics became available [48,49]. Huang et al. [50] subsequently developed an activity-based provincial framework that links national sectoral baselines with provincial statistics and temporal proxies. However, the provincial Carbon Monitor framework reports emissions across only five broad sectors, limiting its ability to distinguish heterogeneous emission changes within industrial subsectors and support more targeted mitigation assessments [51]. Previous studies have demonstrated substantial differences in emission characteristics and mitigation opportunities across industrial subsectors, particularly in energy-intensive industries such as steel, aluminum, and cement [52,53]. Furthermore, fixed activity-to-emission relationships may not fully capture changes in fuel composition, energy efficiency, and technology [10]. Although near-real-time approaches are designed to track recent emission dynamics rather than reproduce the full sectoral detail of conventional inventories, integrating their temporal advantages with more detailed sectoral information could provide a better understanding of recent industrial emission changes.

Here, we update the established provincial framework to provide daily, five-sector $CO_2$ estimates for 31 provincial-level administrative units in mainland China from 1 January 2019 to 31 December 2025. We further disaggregate monthly provincial industrial $CO_2$ emissions into 40 industrial and construction subsectors across 30 provinces, combining historical CEADs subsectoral emission inventories with monthly industrial production indicators while preserving each province's monthly industrial emission total. We examine national and sectoral trajectories, provincial emission levels and recent changes, and persistence across the six annual transitions. Two equal three-year windows, 2019-2022 and 2022-2025, are used to compare provincial growth rates. We further retain the province-sector matrix to quantify gross increases, gross reductions, and their cancellation in the national balance, and evaluate the estimates against CEADs over the overlapping years. Together, these analyses locate recent changes, distinguish slowing growth from actual reductions, and identify what a small national net change conceals. By combining near-real-time emission tracking with finer sectoral detail, this study provides a more comprehensive basis for understanding regional and sectoral emission dynamics and supporting more differentiated mitigation assessments and planning.

## 2 Methods

We estimated daily $CO_2$ emissions from fossil-fuel combustion and industrial processes in mainland China during 2019–2025. The spatial domain covers 31 provincial-level administrative units in mainland China, excluding Hong Kong, Macao, and Taiwan. $CO_2$ emissions were estimated for five major sectors, covering power generation, industry, ground transport, residential activities, and domestic aviation. Provincial emissions were estimated using sector-specific energy and activity statistics together with monthly and daily activity indicators. The detailed methods are described below, and data sources and spatial coverage are documented in **Supplementary Table S1**.

### 2.1 Provincial near-real-time $CO_2$ emissions

Provincial daily emissions were constructed following Huang et al. [50], using national sectoral emissions from China's 2021 National Greenhouse Gas Inventory (NGHGI) as the reference baseline [54]. Sector-specific activity data were used to allocate national emissions across provinces, while high-frequency activity indicators were used to characterize temporal variations. Emissions for other years were estimated by scaling the reference emissions according to changes in sector-specific activity indicators, assuming a constant emission-to-activity ratio:

$$E_{p,s,t} = E_{p,s,b} \times \frac{A_{p,s,t}}{A_{p,s,b}} \quad (1)$$

where $E$ denotes $CO_2$ emissions, $A$ is the sector-specific activity indicator, and $p$ and $s$ denote province and sector. The indices $t$ and $b$ denote the target and reference periods, respectively, with $b$ corresponding to the year 2021. In the following equations, $N$ denotes the national total, $q$ indexes provinces, $y$ denotes year, $m$ denotes month, and $d$ and $k$ index days or months as specified.

This activity-based approach captures emission changes associated with the selected indicators without explicitly accounting for changes in emission factors over time. Spatial allocation and temporal disaggregation are performed using sector-specific data at the available temporal scales, with normalized daily profiles preserving the corresponding monthly or annual emission totals. The detailed estimation procedures for each sector are described below.

**Power generation**

For power generation, the 2021 national baseline was distributed across provincial months in proportion to thermal-power generation reported by the National Bureau of Statistics (NBS). Summation over all provinces and all months in the baseline year normalizes the allocation:

$$E_{p,pow,m} = E_{N,pow,b} \times \frac{TP_{p,m}}{\sum_q \sum_{k \in b} TP_{q,k}} \quad (2)$$

where $TP$ is thermal-power generation and the denominator is total national generation during the 2021 baseline year. Daily coal consumption at major power plants provides the within-month allocation:

$$E_{p,pow,d} = E_{p,pow,m} \times \frac{CC_{p,d}}{\sum_{k \in m} CC_{p,k}} \quad (3)$$

Here, $CC$ denotes daily steam-coal consumption at major power plants, reported by the China Coal Transport and Distribution Association (CCTD) for 25 provinces. The mean daily profile of these provinces was used for the remaining six provinces. Each profile was normalized within its month, preserving the monthly total derived from thermal-power generation. Emissions are attributed to the generating province.

The monthly thermal-power series anchors the geographical allocation and monthly magnitude, whereas daily coal consumption determines the within-month profile. Applying an average normalized profile to provinces without direct coal observations retains the assigned monthly total but provides a less locally constrained representation of daily variability. Changes in electricity supplied to other provinces remain attributed to the location of generation.

**Industry**

The industrial estimate uses crude-steel and cement production as indicators for the temporal evolution of industrial emissions. The NGHGI adopts the source classification of the Intergovernmental Panel on Climate Change (IPCC). Its fuel-combustion categories for iron and steel (1.A.2.a) and non-metallic minerals (1.A.2.f) provide the respective baselines in the supplied framework [50,54]. Provincial monthly production reported by the NBS determines the allocation:

$$E_{p,i,m} = E_{N,i,b} \times \frac{P_{p,i,m}}{\sum_q \sum_{k \in b} P_{q,i,k}}, \ i \in \{steel, cement\} \tag{4}$$

where $P$ denotes physical production and $i$ identifies crude steel or cement. The denominator is national production of product $i$ in the baseline year. Daily production indicators from the Kayrros Industrial Activity Monitor, a satellite-based service for tracking industrial production, were used to redistribute each monthly total:

$$E_{p,i,d} = E_{p,i,m} \times \frac{K_{p,i,d}}{\sum_{k \in m} K_{p,i,k}} \tag{5}$$

Here, $K$ denotes the daily production indicator. Steel and cement jointly account for 66% of baseline energy-related industrial emissions. Their combined daily emissions were scaled by the inverse of this share to estimate the industrial total:

$$E_{p,ind,d} = \frac{E_{p,steel,d} + E_{p,cement,d}}{r_{sc}}, \ r_{sc} = 0.66 \tag{6}$$

This extrapolation assumes that other industrial sources share the combined temporal evolution of steel and cement. For provinces without direct daily production proxies, the mean normalized profile from provinces with observations was used to distribute the monthly total. Purchased electricity is accounted for in power generation rather than counted again in industry. The industrial estimate thus resolves the aggregate trajectory implied by the production indicators; it does not distinguish changes in output from changes in fuel mix or energy use per unit of output.

## Ground transport

National annual ground-transport emissions were distributed among provinces in proportion to civil-vehicle ownership reported by the NBS. This statistic sets the provincial share of the annual total, while the congestion-based indicator below determines its daily distribution:

$$E_{p,tr,y} = E_{N,tr,y} \times \frac{V_{p,y}}{\sum_q V_{q,y}} \tag{7}$$

where $V$ denotes civil-vehicle ownership and the denominator sums all provinces. Daily traffic variation was represented by congestion delay indices from Amap (Gaode Maps), a digital mapping and navigation platform that provides urban traffic information. The congestion index $X$ was converted to a traffic-flow proxy $F$ using

$$F_{c,d} = \alpha + \beta \frac{X_{c,d}{}^{\gamma}}{\delta^{\gamma} + X_{c,d}{}^{\gamma}} \tag{8}$$

where $c$ denotes a monitored urban area, $\alpha = 11{,}089.30$, $\beta = 23{,}460.82$, $\gamma = 1.40$, and $\delta = 17.93$. These parameters were calibrated from daily traffic-flow and congestion observations in Beijing, following Huang et al. [50]. Urban congestion proxies were aggregated using population weights:

$$F_{p,d} = \sum_{c \in p} POP_c F_{c,d} \tag{9}$$

where $POP_c$ is the population of monitored urban area $c$, and the sum includes monitored urban areas within province $p$. Daily provincial emissions were obtained by normalizing the resulting traffic proxy over the calendar year:

$$E_{p,tr,d} = E_{p,tr,y} \times \frac{F_{p,d}}{\sum_{k \in y} F_{p,k}} \tag{10}$$

This normalization preserves the annual provincial total. For provinces without congestion observations, the mean normalized daily profile of provinces with observations was applied. Vehicle ownership and congestion represent different dimensions of transport activity. Ownership determines the spatial allocation, whereas congestion provides the temporal profile for monitored urban areas. Population weighting gives greater influence to observations from more populous areas, but it does not provide direct measurement of all provincial passenger and freight fuel consumption. The normalization ensures that uncertainty in the daily profile is distinct from uncertainty in the annual provincial allocation.

**Domestic aviation**

Domestic aviation emissions were allocated to departure provinces using weekly domestic flight counts from VariFlight. A constant emission factor per flight was calculated by dividing national aviation emissions in the 2021 baseline year by the corresponding total number of domestic departures:

$$EF_{avi} = \frac{E_{N,avi,b}}{\sum_{p,d \in b} FL_{p,d}} \tag{11}$$

Daily aviation emissions for each province were then estimated as:

$$E_{p,avi,d} = EF_{avi} \times FL_{p,d} \tag{12}$$

where $FL$ denotes the daily number of domestic departures. $EF_{avi}$ was held constant when extending the estimates beyond the baseline year, so daily emissions vary in proportion to flight activity. This approach does not explicitly account for differences in aircraft type, route length, passenger load, or operational efficiency.

**Residential activities**

Residential emissions were calculated separately for cooking and space heating, following Huang et al. [50]. The baseline corresponds to the IPCC "Other sectors" category (1.A.4), which includes residential, commercial and institutional, and agricultural activities. Liquid and gaseous fuels were assigned primarily to cooking and water heating, and solid fuels to space heating. Annual provincial supplies of natural gas ($NG$), coal gas ($CG$), and liquefied petroleum gas ($LPG$) from the NBS were combined into the cooking allocation indicator:

$$G_{p,y} = NG_{p,y} + CG_{p,y} + LPG_{p,y} \tag{13}$$

The national annual cooking estimate was allocated among provinces using $G$. Monthly variability was represented by population-weighted night-time lights (NTL) from the Visible Infrared Imaging Radiometer Suite (VIIRS). Observations were screened by valid-pixel counts, and missing values were imputed using a dynamic factor model that exploits interprovincial correlations [50] The monthly indicator was normalized within each year:

$$E_{p,cook,m} = E_{N,cook,y} \times \frac{G_{p,y}}{\sum_q G_{q,y}} \times \frac{NTL_{p,m}}{\sum_{k \in y} NTL_{p,k}} \tag{14}$$

where $NTL_{p,m}$ is the monthly provincial indicator. Cooking emissions were distributed uniformly within each calendar month:

$$E_{p,cook,d} = \frac{E_{p,cook,m}}{D_m} \tag{15}$$

where $D_m$ is the number of days in month $m$. Space-heating emissions were allocated to days using population-weighted heating-degree days (HDD), derived from daily 2 m air temperature in ERA5, the fifth-generation atmospheric reanalysis of the European Centre for Medium-Range Weather Forecasts. Population weighting used the 100 m dataset based on China's seventh national census [55]. ERA5 is described by Hersbach et al. [56]:

$$E_{p,heat,d} = E_{N,heat,y} \times \frac{HDD_{p,d}}{\sum_{q,k \in y} HDD_{q,k}} \tag{16}$$

The denominator sums $HDD$ across provinces and days in year $y$. Daily cooking and heating estimates were combined as

$$E_{p,res,d} = E_{p,cook,d} + E_{p,heat,d} \tag{17}$$

Cooking and heating are constrained at different temporal scales. The cooking profile varies monthly with the night-time-light indicator and is uniform within a month, whereas heating responds to daily population-weighted temperature demand. Daily resolution therefore has different information content for these two components. The residential label retains the broader NGHGI category 1.A.4 coverage and should not be interpreted as a separately measured household-only inventory.

The same activity-based scaling and normalized allocation procedures were applied consistently over 2019-2025. Provincial estimates were summed to national sectoral totals and then across the five sectors:

$$E_{N,s,t} = \sum_{p=1}^{31} E_{p,s,t}; \quad E_{N,t} = \sum_{s \in S} E_{N,s,t} \tag{18}$$

## 2.2 Further sectoral disaggregation of Carbon Monitor $CO_2$ emissions

The five-sector Carbon Monitor framework enables near-real-time tracking of provincial $CO_2$ emissions, but its broad sectoral classification limits the identification of specific emission sources and the assessment of targeted mitigation priorities. The more detailed sectoral classification of CEADs provides a basis for further disaggregation of Carbon Monitor emissions, although the sectoral categories of the two inventories cannot always be directly matched. Further disaggregation is therefore constrained by differences in sectoral definitions and the availability of corresponding annual emission data and monthly activity indicators. For example, CEADs combines different modes of transportation with storage, postal, and telecommunication services into a single category, whereas Carbon Monitor distinguishes ground transport and domestic aviation. This mismatch prevents direct alignment of the two classifications without additional allocation assumptions. For the industrial sector, CEADs provides detailed annual subsectoral emission inventories [9,27,28], while monthly industrial production statistics provide corresponding activity indicators. We therefore restricted further disaggregation to the industrial sector, retaining the original classification of the other four Carbon Monitor sectors.

The provincial industrial $CO_2$ emissions estimated in **Section 2.1** were disaggregated into 40 industrial and construction subsectors for 2019–2025, following the CEADs classification. These subsectors include iron and steel, non-metallic mineral products, chemicals, non-ferrous metals, and construction. Production and supply of electric power, steam, and hot water were excluded to avoid overlap with the Carbon Monitor power sector. The disaggregation combined annual subsectoral energy-related direct $CO_2$ emissions from CEADs with monthly industrial production indicators to derive monthly subsectoral emission allocation weights, while preserving the provincial Carbon Monitor industrial emission totals for each month. Because the CEADs energy-related direct $CO_2$ emission data used in this study were available only through 2022, different approaches were adopted for 2019–2022 and 2023–2025. The disaggregation was implemented for 30 provinces. Xizang was excluded due to the unavailability of the required CEADs industrial subsector emission data, while its emissions remained included in the original five-sector Carbon Monitor inventory. CEADs data were used to derive the historical subsectoral allocation weights rather than to estimate or adjust the aggregate Carbon Monitor industrial emissions. The fuel-specific emission parameters underlying the CEADs inventories are summarized in **Supplementary Table S2** for provenance; these parameters were not directly applied in the activity-based allocation conducted in this study.

For 2019–2022, annual baseline weights were derived from the corresponding CEADs subsectoral $CO_2$ emission inventories. For subsector $i$ in province $p$ and year $y$, the energy-related direct $CO_2$ emissions, $C_{piy}$, were calculated as the sum of emissions from 17 fossil-fuel categories, excluding industrial process emissions. The annual baseline weight, $w_{piy}$, was then calculated by normalizing these emissions across the 40 retained subsectors:

$$w_{piy} = \frac{C_{piy}}{\sum_{j=1}^{40} C_{pjy}} \tag{19}$$

For 2023–2024, annual baseline weights were extrapolated from the preceding year's final annual subsector emission shares using year-on-year changes in the annual output of representative industrial products. The updated weights were normalized across the 40 subsectors within each province:

$$w_{piy} = \frac{S_{pi,y-1} g_{piy}}{\sum_{j=1}^{40} S_{pj,y-1} g_{pjy}} \tag{20}$$

Here, $S_{pi,y-1}$ denotes the final annual emission share of subsector $i$ in the preceding year, calculated as the ratio of its annual emissions, summed from monthly estimates, to the corresponding provincial annual industrial emissions. The growth factor $g_{piy}$ represents the ratio of representative-product output in year $y$ to that in year $y-1$. When comparable annual production data were unavailable, or when either year's product series had been excluded following quality assessment, $g_{piy}$ was set to 1.

For 2019–2024, monthly activity profiles were derived from the normalized monthly output of representative industrial products. When suitable product data were unavailable or failed quality checks, the corresponding provincial Carbon Monitor industrial emission profile was used instead. The representative products assigned to each industrial subsector are listed in **Supplementary Table S3**, and the data quality-control procedures are described in **Supplementary Note 1**. The monthly activity profile, $a_{piym}$, was calculated as follows:

$$a_{piym} = \frac{Q_{piym}}{\sum_{k=1}^{12} Q_{piyk}} \quad \text{or} \quad \frac{E_{pym}^{ind}}{\sum_{k=1}^{12} E_{pyk}^{ind}} \tag{21}$$

where $Q_{piym}$ denotes the monthly output of the representative product, and $E_{pym}^{ind}$ denotes the provincial monthly Carbon Monitor industrial $CO_2$ emissions. The annual baseline weights were combined with the monthly activity profiles and normalized across the 40 subsectors to obtain the monthly allocation shares, $s_{piym}$. Monthly subsectoral $CO_2$ emissions, $E_{piym}$, were then calculated by multiplying these shares by the corresponding provincial Carbon Monitor industrial emissions, as follows:

$$E_{piym} = E_{pym}^{ind} s_{piym}, \quad s_{piym} = \frac{w_{piy} a_{piym}}{\sum_{j=1}^{40} w_{pjy} a_{pjym}} \tag{22}$$

For 2025, the final monthly allocation shares from the corresponding months of 2024 were applied to the 2025 provincial Carbon Monitor industrial emissions, as follows:

$$E_{pi,2025,m} = E^{ind}_{p,2025,m} s_{pi,2024,m} \quad (23)$$

This approach preserves the provincial Carbon Monitor industrial emission total for each month. Final annual subsectoral emission shares are calculated from the resulting monthly estimates and are not constrained to reproduce the annual baseline weights. For 2025, the monthly allocation shares remain fixed at their 2024 values, although annual shares may change with the seasonal distribution of provincial industrial emissions. The estimates reflect emission variations inferred from industrial activity indicators and the adopted allocation assumptions, without explicitly accounting for changes in fuel composition or emission intensity per unit of output. For subsequent analyses, the 40 industrial and construction subsectors were further aggregated into five categories, namely energy production, heavy manufacturing, light manufacturing, high-tech industry and construction, following the original classification of Shan et al. [9]. The mapping of the 40 subsectors to these five categories is provided in **Supplementary Table S4**.

### 2.3 Analysis of emission trajectories and contributions

Daily provincial $CO_2$ emissions were aggregated to monthly and annual totals, and national emissions were calculated by summing across the 31 provinces and five sectors. We analyzed annual emission levels, sectoral shares, year-on-year changes, and cumulative changes during 2019–2025. A centered 12-month moving average was applied to the monthly national emission series to characterize longer-term trends.

Provincial emission trajectories were assessed using the direction of annual changes and compound annual growth rates (CAGRs) over two equal three-year periods, 2019–2022 and 2022–2025. The CAGR for province $q$ between years $y_0$ and $y_1$ was calculated as

$$CAGR(q, y_0 - y_1) = \left(\frac{E(q, y_1)}{E(q, y_0)}\right)^{\frac{1}{y_1 - y_0}} - 1 \quad (24)$$

where $E(q, y)$ denotes annual provincial $CO_2$ emissions. The difference between the two CAGRs, expressed in percentage points per year, was used to assess changes in provincial emission growth rates. A negative difference indicates a lower growth rate in the later period, which may reflect slower growth or a shift towards decline.

To quantify the offsetting emission changes underlying the national balance, we calculated emission differences between 2024 and 2025 for each province–sector combination. Positive changes were summed to obtain gross emission increases, while the absolute values of negative changes were summed to obtain gross reductions. The difference between gross increases and reductions represents the national net emission change.

### 2.4 Evaluation against independent inventories

To contextualize the national emission estimates, we compared annual Carbon Monitor totals with six existing inventories, including CEADs, MEIC, EDGAR, the Global Carbon Budget (GCB), the Energy Institute (EI), and NGHGI [29,54,57]. The comparison covered overlapping years during 2019–2025. Where possible, emissions were selected to approximate the Carbon Monitor accounting scope, while published totals were retained when consistent sectoral alignment was not feasible. NGHGI estimates for 2020 and 2021 were derived from fossil-fuel combustion and cement-process $CO_2$ emissions. Differences in sectoral and geographical coverage were considered when interpreting the comparisons.

We also evaluated our provincial $CO_2$ emission estimates against CEADs for the overlapping period 2019–2022 [9,27,28]. The comparison was conducted at the provincial annual level, with consistent temporal coverage, emission units, and accounting boundaries for fossil-fuel combustion and cement-process emissions. Provincial totals were obtained by aggregating emissions across the respective sectors of each inventory, without imposing a one-to-one sectoral mapping. This total-level comparison was adopted because the sectoral classifications of CEADs and Carbon Monitor do not directly correspond, particularly for transport and electricity-related activities. As CEADs covers 30 mainland provinces, Xizang was excluded from both datasets, yielding 120 province–year pairs.

For each province-year pair, we calculated the ratio of Carbon Monitor to CEADs emissions and the corresponding signed and absolute relative differences, using CEADs as the reference. The median, interquartile range, and proportions of estimates within ±10% and ±20% were used to characterize the distribution of differences. We also compared aggregate emissions across the overlapping provinces and years and calculated Pearson's correlation coefficient and $R^2$ to assess correspondence between the two inventories.

This evaluation concerns the original five-sector Carbon Monitor estimates. Although CEADs is compiled separately, the two inventories may share official activity statistics. The comparison therefore assesses inter-inventory consistency rather than providing fully independent observational validation. The uncertainty assumptions and estimates reported by the original framework [50] are summarized in **Section 3.4**.

# 3 Results

## 3.1 National and sectoral $CO_2$ emission trends during 2019–2025

National $CO_2$ emissions increased from 10.83 Gt $CO_2$ in 2019 to a peak of 12.15 Gt $CO_2$ in 2023, before declining slightly to 12.11 Gt $CO_2$ in 2024 and 12.08 Gt $CO_2$ in 2025 (**Figure 1**). Emissions in 2025 were 0.23% lower than in 2024 but remained 11.6% above the 2019 level. Monthly emissions in 2025 ranged from 927 Mt $CO_2$ in April to 1,120 Mt $CO_2$ in December.

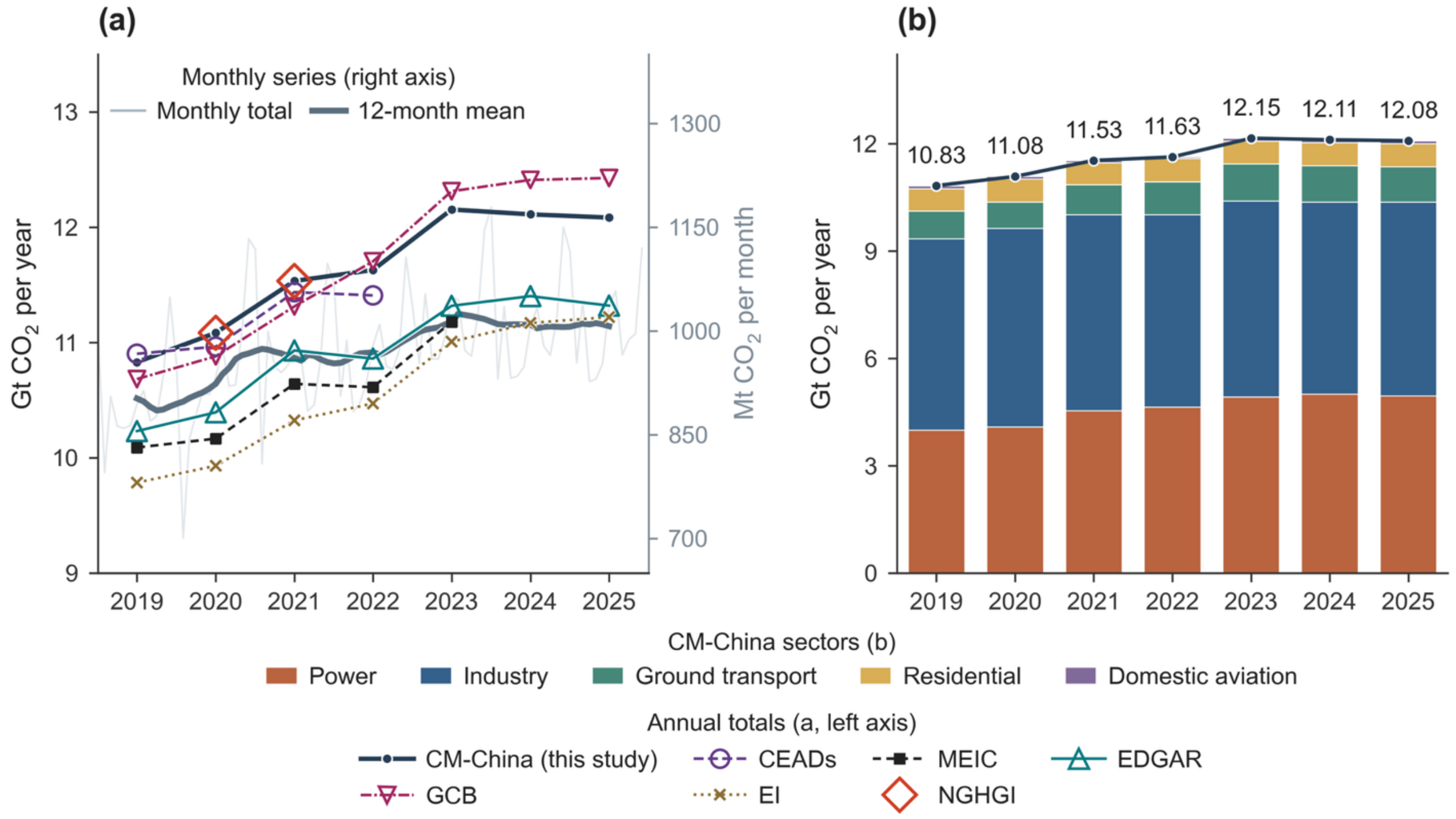

**Figure 1.** National $CO_2$ emissions in China from 2019 to 2025. (a) Monthly $CO_2$ emissions and the 12-month moving mean estimated by CM-China (right axis), together with annual emission estimates from CM-China, CEADs, MEIC, EDGAR, the Global Carbon Budget (GCB), the Energy Institute (EI), and NGHGI (left axis). (b) Annual CM-China $CO_2$ emissions by five major sectors, shown as stacked bars, with national totals indicated above each bar. CEADs estimates

cover 30 provinces, excluding Xizang, while NGHGI estimates for 2020 and 2021 include $CO_2$ emissions from fossil-fuel combustion and cement production.

Industry and power generation dominated national emissions in 2025, contributing 5.42 Gt $CO_2$ (44.8%) and 4.95 Gt $CO_2$ (41.0%), respectively. Ground transport accounted for 8.2%, followed by residential activities (5.3%) and domestic aviation (0.7%). Between 2024 and 2025, industrial emissions increased by 48.8 Mt $CO_2$, nearly offsetting a 46.6 Mt reduction in power generation. Ground transport emissions declined by 27.7 Mt $CO_2$, while residential emissions decreased by 2.6 Mt and domestic aviation emissions increased by 0.35 Mt. The increase in industrial emissions largely offset the reduction in power generation, leaving the decline in ground transport emissions as the main contributor to the small national net reduction of 27.7 Mt $CO_2$.

Comparison with existing inventories showed that CM-China emissions differed by −0.9% to +1.7% from CEADs across the same 30 provinces during 2019–2022, and by −2.8% to +2.0% from GCB during 2019–2025 (**Figure 1a**). In contrast, CM-China estimates were 7.3–9.6% higher than MEIC during 2019–2023, and 5.5–7.4% and 7.7–11.7% higher than EDGAR and EI, respectively, during 2019–2025. GCB and EI continued to increase in 2024–2025, whereas CM-China emissions declined slightly after 2023. The 2021 NGHGI estimate was consistent with the baseline used in this study. Differences in geographical coverage, sectoral boundaries and accounting methods limit direct comparability.

### 3.2 Provincial $CO_2$ emission patterns and changes

The three largest emitters in 2025 were Shandong (1.13 Gt $CO_2$), Jiangsu (1.00 Gt $CO_2$), and Hebei (0.94 Gt $CO_2$), together accounting for 25.4% of the national total (**Figure 2a**). However, provincial emission changes showed a different spatial pattern. Between 2024 and 2025, emissions increased in 16 provinces and decreased in 15 (**Figure 2b**). Hebei recorded the largest absolute reduction (−36.2 Mt $CO_2$), with industry accounting for 82.3% of the decrease, while Zhejiang had the largest increase (+23.1 Mt $CO_2$). In contrast, Gansu experienced a smaller absolute increase (+14.3 Mt $CO_2$) but a higher relative growth rate than Zhejiang (+7.9% versus

+3.9%), highlighting differences between provincial contributions to emission changes and their proportional growth (**Figures 2b and 4a**).

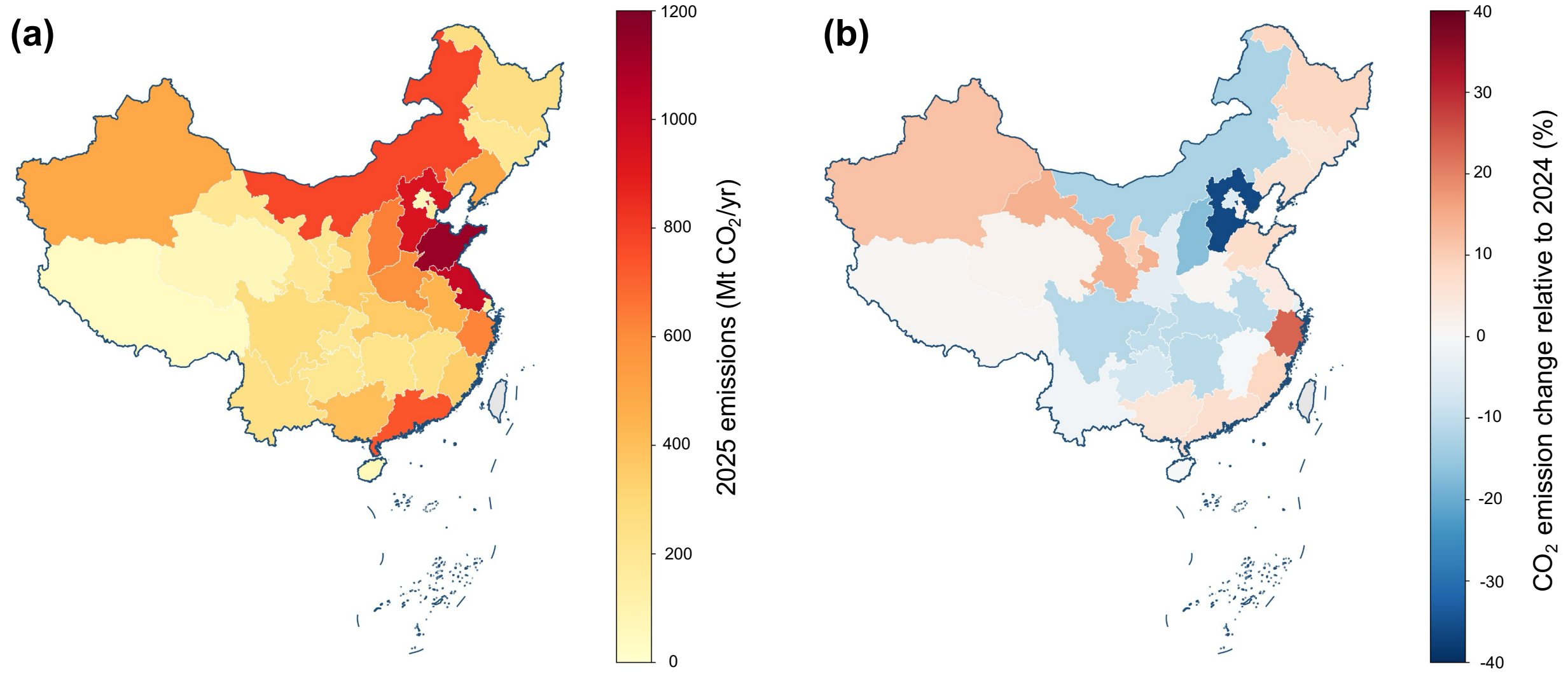

**Figure 2.** Provincial distribution of $CO_2$ emissions. (a) Total emissions in 2025. (b) Percentage change relative to 2024. Grey areas fall outside the accounting domain. Hong Kong, Macao and Taiwan are excluded.

The 2025 sectoral breakdown revealed substantial differences in the sources underlying provincial emission totals (see **Table 1**). Industry was the largest contributor in both Shandong and Hebei, accounting for 608.56 Mt $CO_2$ (53.7%) and 513.01 Mt $CO_2$ (54.7%), respectively, with heavy manufacturing contributing 451.68 and 460.23 Mt $CO_2$, respectively. In contrast, power generation dominated emissions in Inner Mongolia, accounting for 562.18 Mt $CO_2$ (72.9%). Beijing exhibited a different profile, with ground transport forming the largest share at 26.77 Mt $CO_2$ (38.9%).

**Table 1.** Provincial $CO_2$ emissions by sector and industrial category in 2025 (Mt $CO_2$). For the 30 provinces with industrial disaggregation, industry total equals the sum of the five industrial categories. The industrial total for Xizang is retained without subsector disaggregation.

| Province | Total | Power | Ground transport | Residential | Domestic aviation | Industry total | Energy production | Heavy manufacturing | Light manufacturing | High-tech industry | Construction |
|---|---|---|---|---|---|---|---|---|---|---|---|
| Beijing | 68.86 | 19.64 | 26.77 | 7.72 | 5.12 | 9.60 | 0.78 | 4.16 | 1.81 | 0.95 | 1.90 |
| Tianjin | 164.57 | 48.26 | 18.37 | 7.16 | 1.17 | 89.63 | 12.37 | 69.28 | 1.33 | 0.51 | 6.14 |
| Hebei | 937.99 | 265.39 | 85.63 | 73.18 | 0.78 | 513.01 | 46.50 | 460.23 | 5.13 | 0.78 | 0.38 |
| Shanxi | 628.64 | 298.73 | 11.32 | 22.71 | 1.34 | 294.54 | 78.52 | 211.67 | 1.28 | 0.12 | 2.96 |
| Inner Mongolia | 770.75 | 562.18 | 7.82 | 21.44 | 2.50 | 176.80 | 48.08 | 120.89 | 2.84 | 0.12 | 4.87 |
| Liaoning | 503.60 | 174.42 | 23.60 | 30.45 | 3.00 | 272.12 | 40.85 | 227.39 | 1.91 | 1.68 | 0.30 |
| Jilin | 195.83 | 102.58 | 12.49 | 19.69 | 1.39 | 59.68 | 7.11 | 48.77 | 2.13 | 0.04 | 1.63 |
| Heilongjiang | 262.05 | 145.40 | 14.39 | 31.72 | 2.08 | 68.45 | 23.09 | 40.38 | 4.40 | 0.12 | 0.46 |
| Shanghai | 166.29 | 57.08 | 27.23 | 5.24 | 4.68 | 72.07 | 11.45 | 54.80 | 2.56 | 0.58 | 2.68 |
| Jiangsu | 999.78 | 363.54 | 106.93 | 46.69 | 3.63 | 478.99 | 23.04 | 435.31 | 13.99 | 5.74 | 0.92 |
| Zhejiang | 621.02 | 250.21 | 69.74 | 20.67 | 4.62 | 275.77 | 40.92 | 164.75 | 27.61 | 24.03 | 18.46 |
| Anhui | 437.97 | 206.10 | 21.78 | 29.27 | 0.91 | 179.92 | 25.74 | 142.21 | 3.09 | 1.77 | 7.11 |
| Fujian | 345.35 | 143.02 | 35.43 | 8.03 | 3.04 | 155.82 | 14.33 | 127.23 | 5.33 | 3.17 | 5.76 |
| Jiangxi | 246.98 | 93.91 | 20.61 | 11.57 | 1.04 | 119.84 | 10.65 | 104.95 | 2.35 | 1.24 | 0.65 |
| Shandong | 1,134.18 | 351.37 | 94.19 | 76.29 | 3.77 | 608.56 | 119.11 | 451.68 | 28.43 | 3.59 | 5.75 |
| Henan | 576.71 | 250.63 | 46.09 | 61.50 | 1.68 | 216.81 | 51.48 | 145.83 | 2.30 | 3.46 | 13.75 |
| Hubei | 352.75 | 126.60 | 16.44 | 16.01 | 2.27 | 191.43 | 13.60 | 155.68 | 9.90 | 0.73 | 11.52 |
| Hunan | 224.61 | 79.21 | 22.67 | 15.53 | 1.93 | 105.27 | 9.64 | 81.24 | 3.90 | 1.10 | 9.39 |
| Guangdong | 732.78 | 283.51 | 128.08 | 16.44 | 8.55 | 296.19 | 48.51 | 220.57 | 14.82 | 7.62 | 4.67 |
| Guangxi | 399.46 | 85.57 | 22.58 | 6.85 | 1.51 | 282.95 | 4.49 | 246.84 | 28.59 | 0.13 | 2.89 |
| Hainan | 58.54 | 21.15 | 4.63 | 1.64 | 2.98 | 28.14 | 11.42 | 7.87 | 4.90 | 0.02 | 3.93 |
| Chongqing | 189.58 | 59.91 | 42.54 | 4.68 | 2.72 | 79.73 | 12.85 | 57.75 | 3.91 | 0.87 | 4.35 |
| Sichuan | 276.04 | 51.09 | 48.10 | 18.78 | 5.27 | 152.81 | 36.05 | 101.50 | 6.68 | 1.77 | 6.81 |
| Guizhou | 205.70 | 101.05 | 8.12 | 6.05 | 1.64 | 88.83 | 30.39 | 51.03 | 1.00 | 0.04 | 6.38 |
| Yunnan | 239.31 | 69.94 | 11.52 | 9.19 | 4.48 | 144.19 | 34.75 | 102.16 | 2.27 | 0.46 | 4.56 |
| Xizang | 37.18 | 0.35 | 19.32 | 7.98 | 0.63 | 8.90 | N/A | N/A | N/A | N/A | N/A |
| Shaanxi | 352.77 | 184.92 | 23.12 | 20.52 | 2.99 | 121.21 | 74.96 | 42.94 | 0.82 | 0.37 | 2.12 |
| Gansu | 195.29 | 104.09 | 6.36 | 11.38 | 1.44 | 72.01 | 12.16 | 57.33 | 0.50 | 0.08 | 1.95 |
| Qinghai | 66.99 | 10.50 | 3.55 | 6.00 | 0.67 | 46.27 | 11.11 | 33.25 | 0.20 | 0.01 | 1.70 |
| Ningxia | 203.43 | 147.30 | 3.13 | 4.30 | 0.69 | 48.01 | 15.86 | 30.39 | 0.52 | 0.28 | 0.97 |
| Xinjiang | 488.62 | 296.84 | 9.96 | 19.64 | 3.98 | 158.21 | 65.59 | 86.35 | 4.12 | 0.07 | 2.08 |

Longer-term provincial emission trajectories revealed differences in both the persistence and pace of emission growth **(see Figure 3)**. Emissions increased in every year from 2019 to 2025 in Gansu, Jiangsu, Xinjiang, Xizang and Zhejiang, whereas Hunan recorded declines in five of the six annual transitions (**Supplementary Figure S1a**). However, sustained emission growth did not necessarily imply accelerating growth. The compound annual growth rate (CAGR) was lower in 2022–2025 than in 2019–2022 in 19 of the 31 provinces, including Zhejiang, where it declined from 5.44% to 3.63% (**Supplementary Figure S1c**). Over the entire period, cumulative emission changes ranged from −12.0% in Guizhou to +35.0% in Xizang (**Supplementary Figure S1b**). Despite recording the highest relative increase, Xizang contributed only 9.6 Mt $CO_2$ to the national emission increase, compared with 167.7 Mt $CO_2$ from Inner Mongolia, illustrating the importance of initial emission levels in determining absolute contributions.

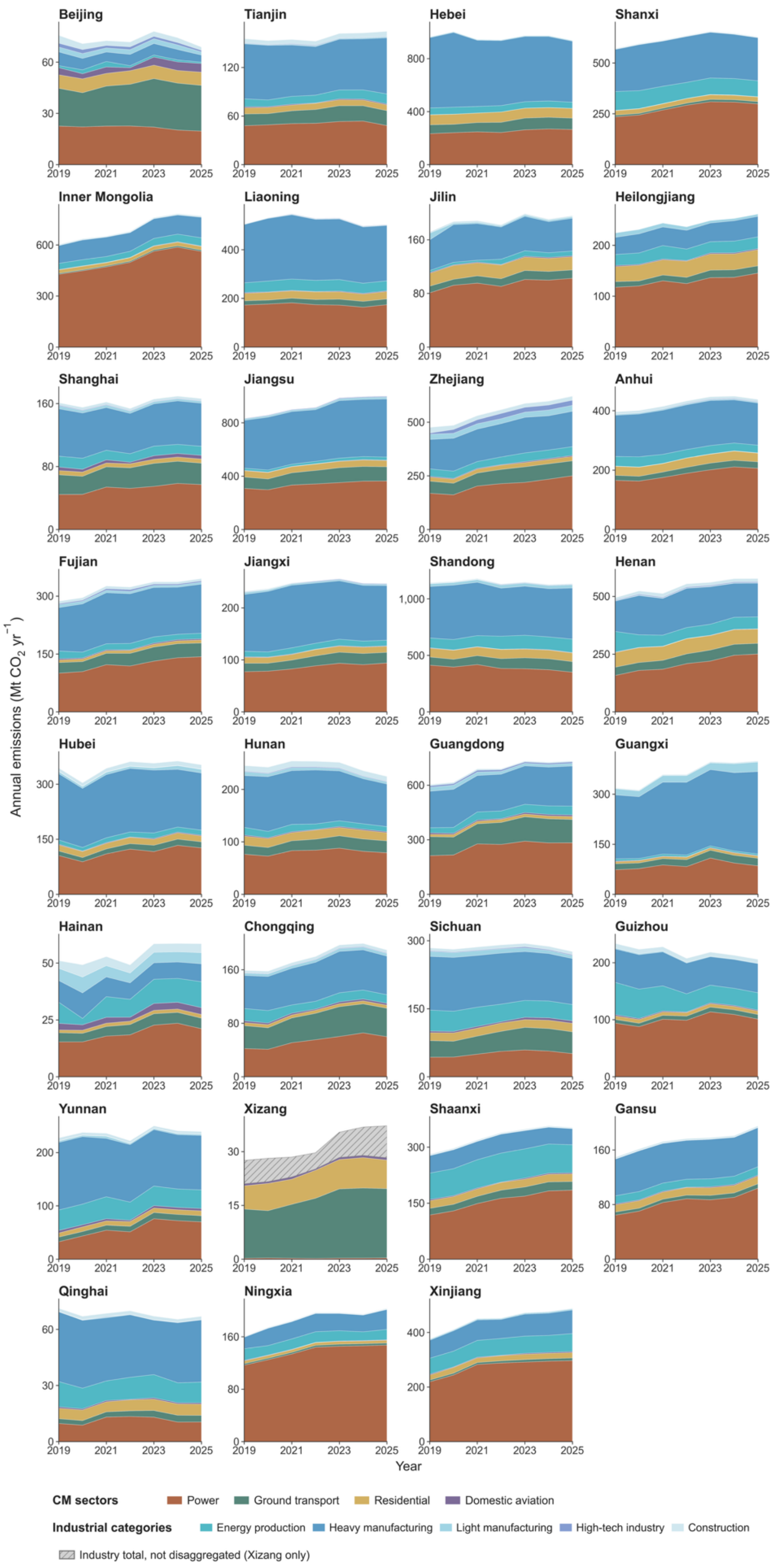

Beijing
Tianjin
Hebei
Shanxi
Inner Mongolia
Liaoning
Jilin
Heilongjiang
Shanghai
Jiangsu
Zhejiang
Anhui
Fujian
Jiangxi
Shandong
Henan
Hubei
Hunan
Guangdong
Guangxi
Hainan
Chongqing
Sichuan
Guizhou
Yunnan
Xizang
Shaanxi
Gansu
Qinghai
Ningxia
Xinjiang
Annual emissions (Mt $CO_2$ $yr^{-1}$)
Year
CM sectors
Power
Ground transport
Residential
Domestic aviation
Industrial categories
Energy production
Heavy manufacturing
Light manufacturing
High-tech industry
Construction
Industry total, not disaggregated (Xizang only)

**Figure 3.** Annual provincial $CO_2$ emissions by sector and industrial category during 2019–2025. Panels use independent y-axis scales. The 2025 industrial-category estimates use corresponding province-month shares from 2024. For Xizang, hatching denotes undisaggregated industry total.

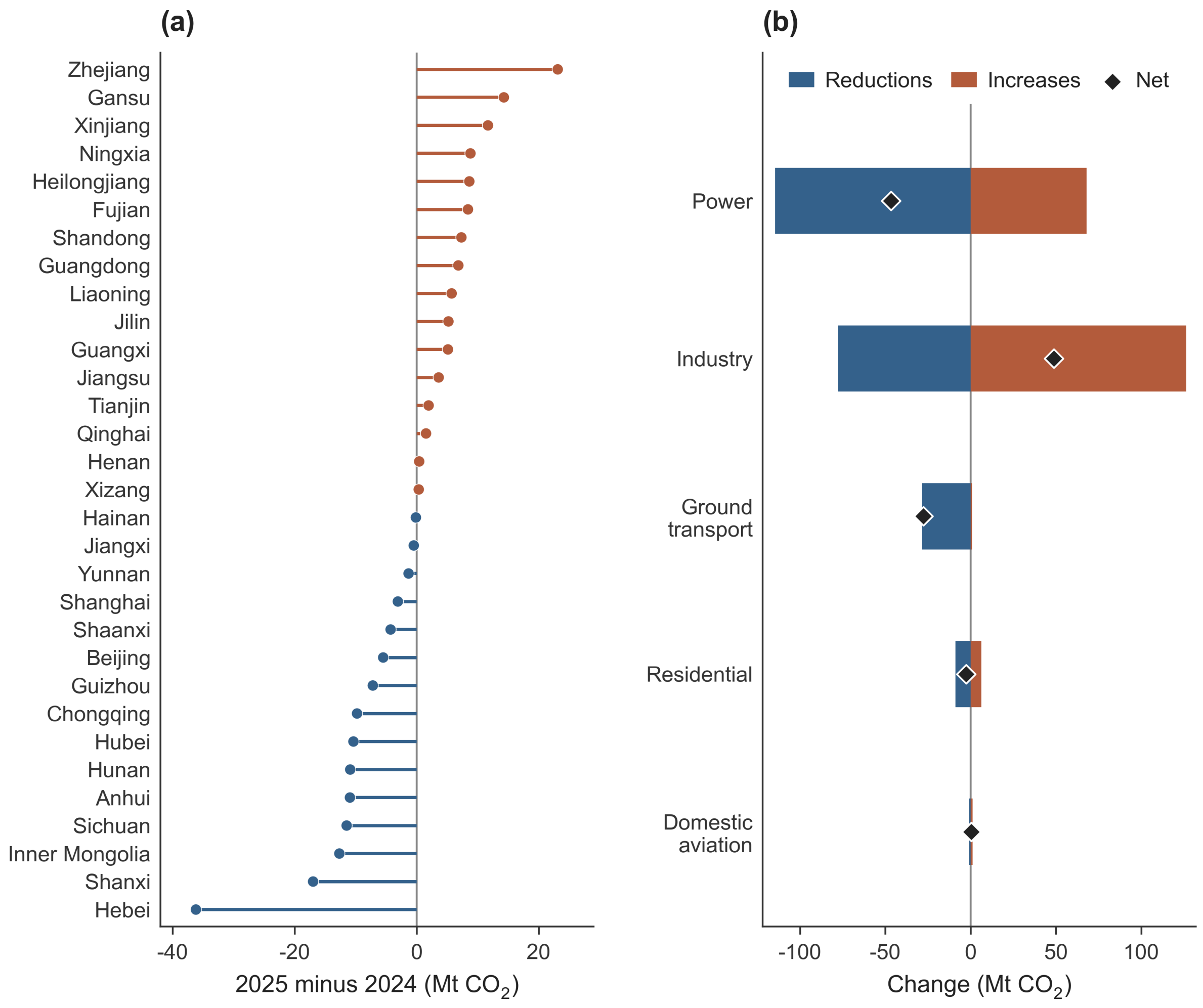

**Figure 4.** Provincial and sectoral contributions to $CO_2$ emission change from 2024 to 2025. (a) Net change across five sectors in each province. (b) Gross reductions and increases summed separately across provinces within each sector. Diamonds mark the national net change for each sector.

Although national $CO_2$ emissions declined by only 27.7 Mt $CO_2$ between 2024 and 2025, this net change reflected substantially larger gross emission increases of 203.0 Mt $CO_2$ and gross reductions of 230.7 Mt $CO_2$ across provinces and sectors (**Figure 4b**). Sectoral contributions to provincial emission changes also showed substantial regional differences (**Figure 4**). Within individual sectors, emission increases in some provinces coincided with reductions in others **(Figure 4b**). Moreover, similar directions of sectoral change could result in contrasting provincial outcomes. In Shandong, industrial emissions increased by 32.7 Mt $CO_2$, exceeding the combined reductions in power generation (−22.0 Mt $CO_2$) and the other three sectors (−3.3 Mt $CO_2$), resulting in a net provincial increase of 7.3 Mt $CO_2$. In Inner Mongolia, industrial emissions also increased (+12.6 Mt $CO_2$), but a larger reduction in power generation (−24.9 Mt $CO_2$) contributed to a net provincial decrease of 12.7 Mt $CO_2$. These contrasting cases demonstrate how differences in the relative contributions of industry and power generation shaped provincial emission changes, even when the directions of sectoral changes were similar.

### 3.3 Spatiotemporal patterns of industrial subsector $CO_2$ emissions

Heavy manufacturing dominated industrial $CO_2$ emissions across the 30 provinces in 2024, contributing 4.05 Gt $CO_2$ (75.6%) of the total 5.36 Gt $CO_2$ (Figure 5a). Energy production accounted for 0.93 Gt $CO_2$ (17.3%), followed by light manufacturing (3.4%), construction (2.5%) and high-tech industry (1.1%). At the subsector level, ferrous-metal smelting and pressing alone contributed 2.78 Gt $CO_2$, accounting for 51.9% of total industrial emissions. Petroleum processing and coking, nonferrous-metal smelting and pressing, and nonmetal mineral products each accounted for 8–9% of total industrial emissions. These results indicate that industrial emissions were concentrated in a small number of subsectors, particularly ferrous-metal smelting and pressing.

The relatively stable share of heavy manufacturing concealed contrasting emission trends among its constituent subsectors. Its contribution remained between 74.9% and 76.7% during 2019–2024, while its absolute emissions increased by 56.1 Mt $CO_2$ (**Figure 5a**). Over the same period, emissions from nonferrous-metal smelting and pressing increased by 102.3 Mt $CO_2$ (31.0%) and those from ferrous-metal smelting and pressing increased by 92.4 Mt $CO_2$ (3.4%), whereas emissions from nonmetal mineral products declined by 108.1 Mt $CO_2$ (20.4%). More recently, between 2023 and 2024, ferrous-metal smelting and pressing and nonmetal mineral products recorded reductions of 76.6 Mt $CO_2$ and 62.6 Mt $CO_2$, respectively. These divergent subsector trends were not apparent from the relatively stable aggregate share of heavy manufacturing. In 2025, total industrial emissions across the 30 provinces reached 5.41 Gt $CO_2$, with subsector estimates based on the corresponding provincial monthly allocation shares from 2024.

Provincial industrial emissions showed substantial differences in both magnitude and composition in 2025 (**Figure 5b**). Heavy manufacturing contributed 460.2 Mt $CO_2$ in Hebei and 435.3 Mt $CO_2$ in Jiangsu, accounting for 89.7% and 90.9% of their respective industrial emissions, compared with the 30-province average of 75.5%. In contrast, energy production was the dominant industrial category in Shaanxi, contributing 75.0 Mt $CO_2$ (61.8%), while heavy manufacturing accounted for 42.9 Mt $CO_2$ (35.4%). These differences indicate considerable spatial heterogeneity in industrial emission composition despite the dominance of heavy manufacturing at the aggregate level.

The estimated monthly emission profiles further revealed differences among industrial categories in 2024, even after accounting for variations in calendar-month length (**Figure 5c**). February recorded the lowest daily mean emissions across all five categories, ranging from 16% below the annual daily mean for energy production to 34% below for high-tech industry. However, peak emissions occurred in different months and varied in magnitude. Heavy manufacturing peaked in June, at 6% above its annual daily mean, whereas light manufacturing peaked in December, at 46% above its annual daily mean. Distinct monthly patterns were also evident within heavy manufacturing. Ferrous-metal smelting and pressing reached 8.20 Mt $CO_2$ per day in June, compared with 7.10 Mt $CO_2$ per day in December. In contrast, nonmetal mineral products peaked in October at 1.33 Mt $CO_2$ per day, while reaching a minimum of 0.57 Mt $CO_2$ per day in February. These monthly patterns reflect differences in the industrial activity indicators and allocation assumptions underlying the subsector estimates.

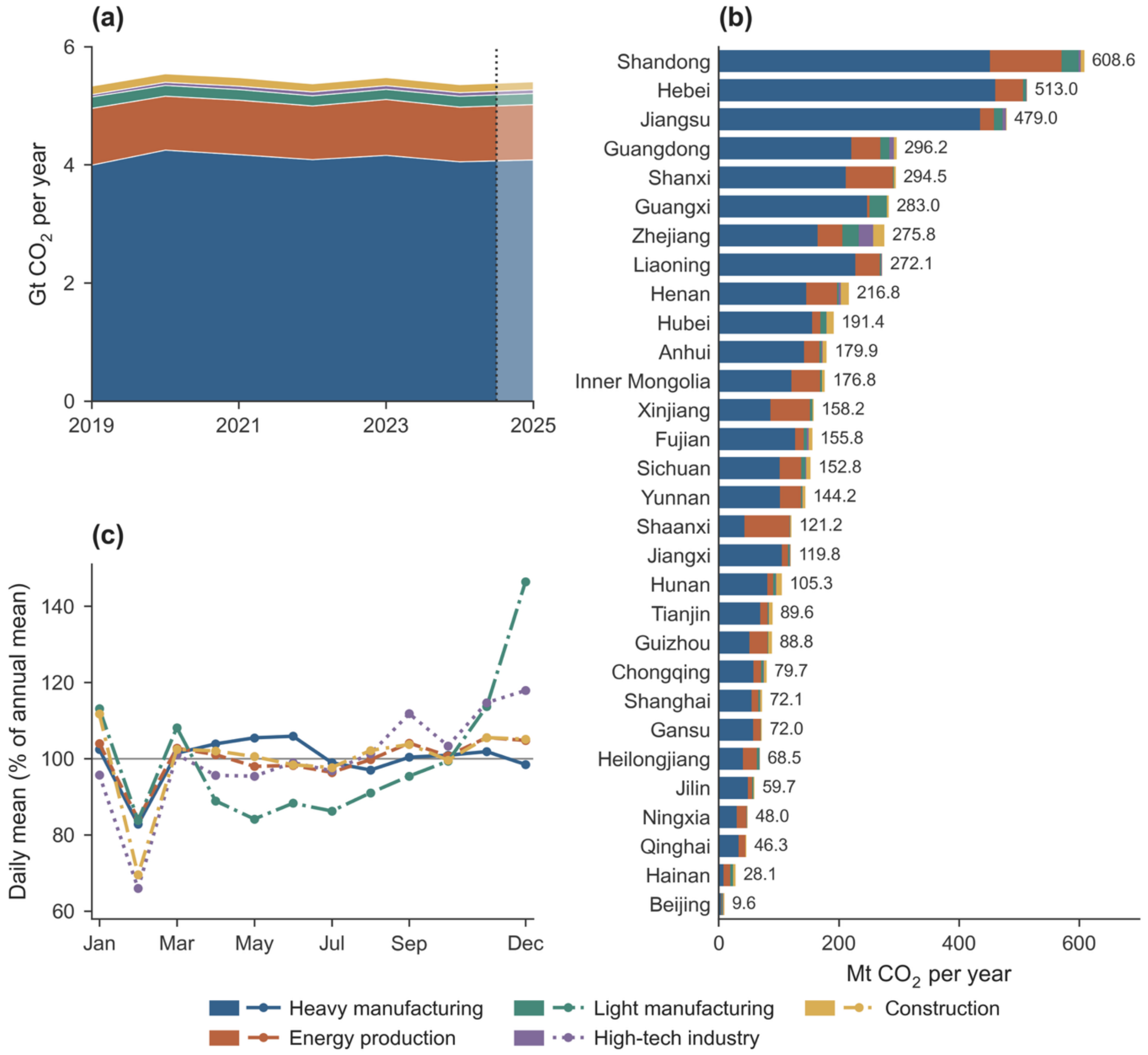

**Figure 5.** Industrial and construction $CO_2$ emissions across 30 mainland Chinese provincial-level regions, excluding Xizang. (a) Annual $CO_2$ emissions by industrial category during 2019–2025. The dotted line and lighter shading distinguish the 2025 estimates, derived using the corresponding province-month allocation shares from 2024. (b) Provincial industrial $CO_2$ emissions by category in 2025, shown as stacked horizontal bars and ordered by total industrial emissions. Values at the end of each bar indicate the provincial industrial emission totals. (c) Monthly mean daily $CO_2$ emissions by industrial category in 2024, expressed relative to the corresponding annual daily mean (100%). The five categories follow the classification adapted from Shan et al. [9]. Energy production excludes electricity, steam and hot-water supply.

### 3.4 Evaluation and uncertainty of emission estimates

Carbon Monitor estimates showed close agreement with CEADs at the aggregate level, although notable differences remained among individual provinces. Across 120 matched province-year observations during 2019–2022, aggregate Carbon Monitor emissions were 0.56% higher than those reported by CEADs, with a Pearson correlation coefficient of 0.971 ($R^2 = 0.942$, **Figure 6a**). The median absolute relative difference was 9.4%, with 52.5% and 80.0% of observations falling within ±10% and ±20% of CEADs, respectively. Signed differences had a median of −1.9% and an interquartile range of −10.0% to +8.7%. Despite the small aggregate difference, systematic provincial deviations persisted. Carbon Monitor estimates were consistently lower than CEADs in Inner Mongolia (−24.2% to −22.7%) and higher in Shandong (+21.7% to +24.5%) across all four years (**Figure 6b**). These results indicate that close agreement in aggregate emissions did not necessarily extend to individual provinces.

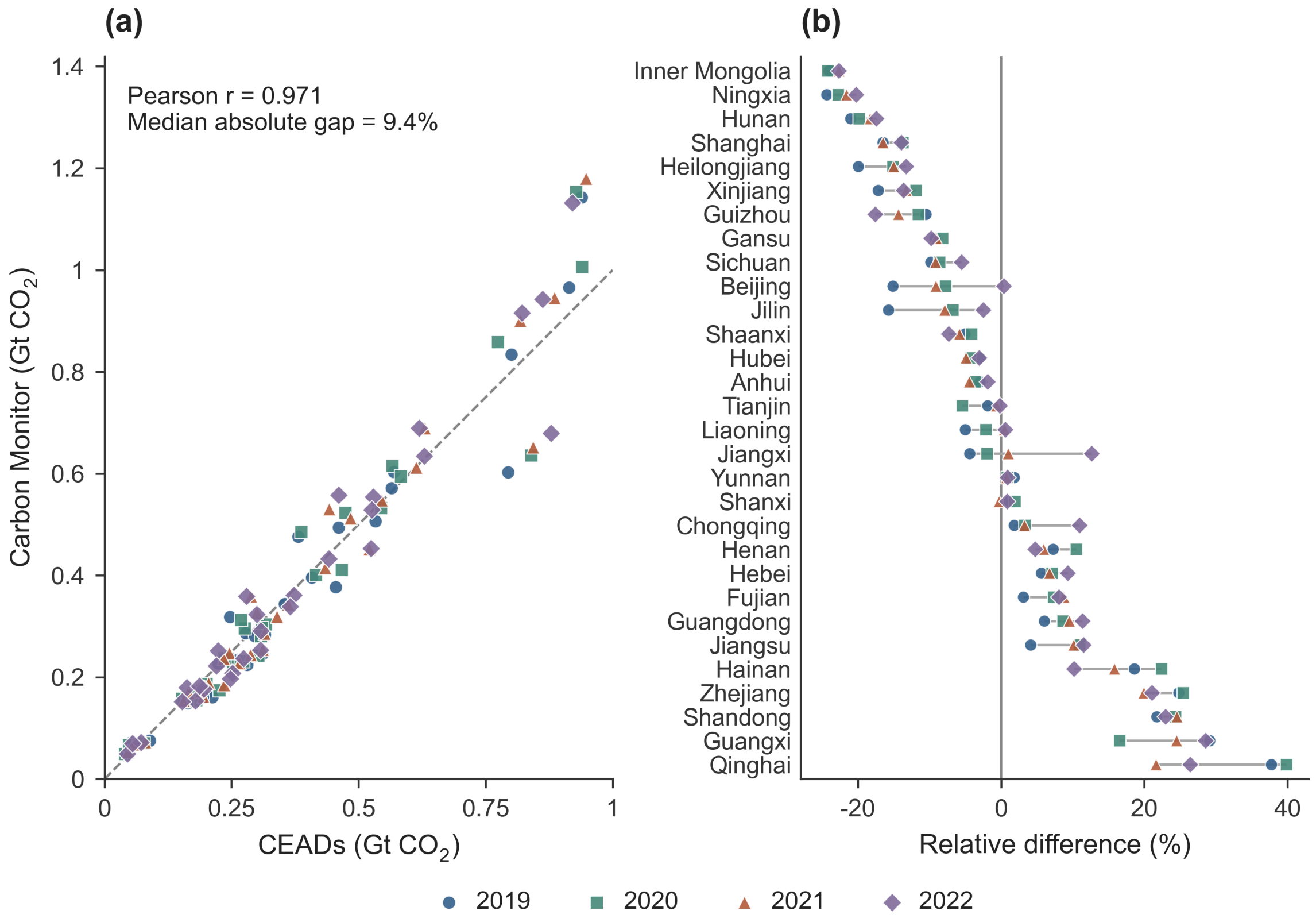

**Figure 6.** Comparison of annual provincial $CO_2$ emissions estimated by Carbon Monitor and CEADs during 2019–2022. (a) Comparison of provincial annual $CO_2$ emissions between the two inventories, with the dashed line indicating the 1:1 relationship. (b) Relative differences in provincial annual $CO_2$ emissions between Carbon Monitor and CEADs, expressed as percentages relative to CEADs and ordered by the provincial median. Horizontal segments indicate the range of annual differences during 2019–2022. Xizang is excluded from the comparison due to the unavailability of CEADs data.

The uncertainty estimates reported for the original provincial emission framework also varied considerably across provinces (**Supplementary Table S5**). The relative uncertainty in provincial total daily $CO_2$ emissions ranged from ±12.9% in Hainan to ±38.6% in Gansu. Power-sector uncertainty was ±23.8% for most provinces but increased to ±73.3% in Beijing, Tianjin, Qinghai, Yunnan and Gansu, where daily activity profiles from other provinces were used to substitute for unavailable proxy data. These estimates characterize uncertainty under the assumptions of the original framework and do not directly quantify uncertainty in recent year-to-year emission changes or the newly disaggregated monthly industrial subsector estimates.

## 4 Discussion

### 4.1 National and provincial emission characteristics

The slight decline in national $CO_2$ emissions after 2023 reflected contrasting sectoral changes rather than uniform reductions. In 2025, national electricity generation increased by 4.8%, while thermal generation decreased by 0.7%. Wind and solar generation increased by 13.1% and 39.8%, respectively [58]. The decline in thermal generation alongside continued growth in renewable electricity generation indicates a shift in China's electricity generation mix. Similarly, the reduction in ground transport emissions coincided with an increase in China's new-energy vehicle stock from 31.4 million at the end of 2024 to 44.0 million at the end of 2025 [58,59]. However, the contributions of vehicle electrification, traffic activity and fuel efficiency cannot be distinguished using the current estimates. Together, these developments provide broader context for the contrasting emission trends across the power and transport sectors. Nevertheless, two consecutive years of small emission decreases are insufficient to establish a sustained national emission peak [8,60].

The heterogeneous provincial trajectories indicate that changes in national emissions do not necessarily represent uniform progress across regions. Their interpretation also depends on the accounting boundary. In particular, the production-based approach attributes power-sector emissions to the province where electricity is generated, rather than where it is consumed [61–63]. Consequently, provincial emission changes should be interpreted alongside regional differences in industrial composition and electricity production, particularly when assessing the geographical distribution of mitigation progress.

### 4.2 Industrial subsector emission characteristics and their implications

The industrial disaggregation provides additional information beyond the five-sector Carbon Monitor framework by revealing contrasting emission trajectories within broad industrial categories. Although the estimated share of heavy manufacturing remained relatively stable during 2019–2024, emissions increased in nonferrous-metal smelting and pressing while decreasing in nonmetal mineral products. These differences indicate that relatively stable aggregate industrial emissions or category shares can conceal substantial changes among individual subsectors. Previous studies have highlighted the importance of production technologies, energy and material inputs, and scrap utilization in determining emissions and mitigation opportunities in China's iron and steel industry [64–66]. While these factors cannot be separately quantified using the present allocation framework, the subsector estimates can help identify industrial activities where more detailed assessments are needed.

Provincial differences in industrial composition and the estimated monthly emission profiles provide further information for assessing regional industrial emission dynamics. Provinces with similar aggregate industrial emissions may have substantially different subsectoral structures, implying different priorities for subsequent emission assessments. Likewise, differences in the timing and magnitude of estimated monthly variations suggest that a common industrial temporal profile may not adequately represent all subsectors. However, these estimates are constrained by provincial industrial emission totals and informed by historical allocation shares and representative production indicators, rather than independent subsector-level emission measurements. High-frequency plant-level monitoring, as demonstrated in China's cement industry [67], could provide complementary evidence to evaluate and refine these estimates.

### 4.3 Limitations and uncertainties

The reliability of the near-real-time provincial emission estimates depends on the representativeness of sector-specific activity indicators and the assumptions used to extrapolate emissions from the 2021 baseline. Changes in fuel composition, energy efficiency, and technology may alter the relationship between activity indicators and actual emissions, while the substitution of missing daily activity indicators limits the representation of province-specific temporal variability. The comparison with CEADs during 2019–2022 assesses inter-inventory consistency rather than fully independent validation, as the two inventories may share underlying activity data. Differences in inventory construction and spatial allocation methods represent additional sources of uncertainty in emission inventories [68–71].

The industrial subsector estimates introduce additional uncertainties associated with representative product indicators and emission allocation assumptions. Individual products may not adequately capture activity changes across entire subsectors, while the use of provincial industrial profiles where product data are unavailable limits the independent characterization of subsector-specific monthly variations. Furthermore, the 2025 estimates retain the corresponding monthly allocation shares from 2024, preventing independent identification of subsequent changes in subsector composition. Independent subsector-level activity and fuel-consumption data are therefore needed to evaluate the disaggregated estimates.

## 5 Conclusion

This study provides near-real-time $CO_2$ emission estimates for 31 provincial-level administrative regions in mainland China during 2019–2025, covering five major sectors. National emissions reached 12.15 Gt $CO_2$ in 2023 before declining slightly to 12.08 Gt $CO_2$ in 2025. Between 2024 and 2025, the national decrease of 27.7 Mt $CO_2$ (0.23%) was accompanied by emission increases in 16 provinces and decreases in 15. These contrasting changes highlight the importance of provincial and sectoral information for interpreting national emission trends.

The further disaggregation of industrial emissions into 40 subsectors across 30 provinces provides additional detail on industrial emission composition and temporal evolution. Heavy manufacturing accounted for 75.6% of industrial emissions in 2024, but its relatively stable share concealed divergent trends among major subsectors. Industrial composition also varied across provinces, while different categories exhibited distinct monthly emission profiles. The 2025 subsector estimates retain the corresponding provincial monthly shares from 2024 and therefore do not independently resolve subsequent structural changes. Overall, the extended dataset provides a more detailed basis for tracking industrial emission changes and identifying priorities for independent evaluation.

## Acknowledgements

None.

## Author contributions

Xiaoting Huang: Conceptualization, Methodology, Visualization, Writing – original draft. Zhu Deng: Conceptualization, Methodology. Ling Ye: Data curation, Visualization. Peng Lu: Data curation. Xuanren Song: Methodology. Yuli Shan: Writing – review & editing. Zhu Liu: Conceptualization, Supervision, Writing – review & editing.

All authors have read and approved the final version of the manuscript.

**Data availability**

The monthly $CO_2$ emission estimates for five sectors across 31 provincial-level regions and for 40 industrial and construction subsectors across 30 provinces are available on Zenodo (https://doi.org/10.5281/zenodo.23998845).

**Declaration of competing interest**

Author 7 is the editor-in-chief of this journal, but he is not involved in the peer-review or decision of this article.

**Funding**

This work is supported by the State Key Laboratory of Hydroscience and Engineering (Project number: sklhse-TD-2025-01).

**Use of AI statement**

During the preparation of this work, the authors used ChatGPT to assist with language polishing and improving the clarity and readability of the manuscript. The authors reviewed and edited the content as needed and take full responsibility for the content of the publication.

# Supplementary Information

China daily $CO_2$ emission estimates 2019–2025

**Table S1.** Data sources and spatial coverage of the provincial estimation framework. Adapted from SI Table 2 of Huang et al. [1].

| Sector | Data source and input | Availability in the supplied SI |
|---|---|---|
| Power | Monthly statistics of thermal-power generation, https://data.stats.gov.cn/english/easyquery.htm?cn=E0101 [Last access: November 10, 2025] | 31 provinces |
| Power | Daily steam coal consumption, https://www.cctd.com.cn/ [Last access: November 10, 2025] | 25 provinces: Shanghai, Jiangsu, Zhejiang, Fujian, Shandong, Guangdong, Guangxi, and Hainan, Hebei, Shanxi, Inner Mongolia, Liaoning, Jilin, Heilongjiang, Anhui, Jiangxi, Henan, Hubei, Hunan, Chongqing, Sichuan, Guizhou, Shaanxi, Ningxia, and Xinjiang |
| Industry | Monthly statistics of cement production and crude steel production, https://data.stats.gov.cn/english/easyquery.htm?cn=E0101 [Last access: November 10, 2025] | 31 provinces |
| Industry | Daily cement and steel production, https://app.kayrros.com/ [Last access: November 10, 2025] | 27 provinces for both cement and steel: Anhui, Chongqing, Fujian, Guangdong, Guangxi, Gansu, Guizhou, Hebei, Heilongjiang, Henan, Hubei, Hunan, Inner Mongolia, Jiangsu, Jiangxi, Jilin, Liaoning, Ningxia, Qinghai, Shaanxi, Shandong, Shanxi, Sichuan, Tibet, Xinjiang, Yunnan, Zhejiang<br>1 province for cement only: Hainan<br>1 province for steel only: Shanghai |
| Ground Transport | Annual provincial civil-vehicle ownership, https://data.stats.gov.cn/english/easyquery.htm?cn=E0103 [Last access: November 10, 2025] | 31 provinces |
| Ground Transport | Daily city congestion index, https://report.amap.com/ [Last access: November 10, 2025] | 99 cities in 30 provinces |
| Aviation | Weekly domestic flight counts by departure province, https://data.variflight.com/RouteAnalysis [Last access: November 10, 2025] | 31 provinces |
| Residential | Annual natural gas, coal gas, and liquefied petroleum gas (LPG) by provinces, https://data.stats.gov.cn/english/easyquery.htm?cn=E0103 [Last access: November 10, 2025] | 31 provinces |
| Residential | Monthly weighted nighttime light, https://eogdata.mines.edu/nighttime_light/monthly/v10/ [Last access: November 10, 2025] | Gridded data with full coverage |
| Residential | Daily 2m temperature, https://cds.climate.copernicus.eu/datasets/reanalysis-era5-single-levels [Last access: November 10, 2025] | Gridded data with full coverage |
| Residential | A 100 m gridded population dataset of China's seventh census, Chen et al. [2] | Gridded data with full coverage |

5

**Table S2.** Fuel-specific emission parameters underlying the CEADs emission inventories. Adapted from Table 1 of Shan et al. [3].

| No. | CEADs fuel category | Fuels in China's energy statistics | $NCV_i$ (PJ per $10^4$ t or $10^8$ $m^3$) | $CC_i$ (t C $TJ^{-1}$) |
|---|---|---|---|---|
| 1 | Raw coal | Raw coal | 0.21 | 26.32 |
| 2 | Cleaned coal | Cleaned coal | 0.26 | 26.32 |
| 3 | Other washed coal | Other washed coal | 0.15 | 26.32 |
| 4 | Briquette | Briquettes; gangue | 0.18 | 26.32 |
| 5 | Coke | Coke | 0.28 | 31.38 |
| 6 | Coke oven gas | Coke oven gas | 1.61 | 21.49 |
| 7 | Other gas | Blast furnace gas; converter gas; other gas | 0.83 | 21.49 |
| 8 | Other coking products | Other coking products | 0.28 | 27.45 |
| 9 | Crude oil | Crude oil | 0.43 | 20.08 |
| 10 | Gasoline | Gasoline | 0.44 | 18.90 |
| 11 | Kerosene | Kerosene | 0.44 | 19.60 |
| 12 | Diesel oil | Diesel oil | 0.43 | 20.20 |
| 13 | Fuel oil | Fuel oil | 0.43 | 21.10 |
| 14 | Other petroleum products | Naphtha; lubricants; paraffin; white spirit; bitumen asphalt; petroleum coke; other petroleum products | 0.51 | 17.20 |
| 15 | LPG | Liquefied petroleum gas (LPG) | 0.47 | 20.00 |
| 16 | Refinery gas | Refinery gas | 0.43 | 20.20 |
| 17 | Natural gas | Natural gas | 3.89 | 15.32 |

**Notes:** $NCV_i$ is net calorific value and $CC_i$ is carbon content. NCV is expressed in PJ per $10^4$ t for solid and liquid fuels and PJ per $10^8$ $m^3$ for gaseous fuels; CC is expressed in t C $TJ^{-1}$. Fuel groupings follow the source table. These parameters document the underlying CEADs inventory and were not directly applied in the activity-based allocation conducted in this study.

**Table S3.** Representative production indicators for the 40 industrial and construction subsectors.

| ID | CEADs subsector | Representative production indicator |
|---|---|---|
| 1 | Coal Mining and Dressing | Raw coal |
| 2 | Petroleum and Natural Gas Extraction | Natural gas |
| 3 | Ferrous Metals Mining and Dressing | Run-of-mine iron ore |
| 4 | Nonferrous Metals Mining and Dressing | None assigned |
| 5 | Nonmetal Minerals Mining and Dressing | Phosphate rock |
| 6 | Other Minerals Mining and Dressing | None assigned |
| 7 | Logging and Transport of Wood and Bamboo | None assigned |
| 8 | Food Processing | Finished sugar |
| 9 | Food Production | None assigned |
| 10 | Beverage Production | Beverages |
| 11 | Tobacco Processing | None assigned |
| 12 | Textile Industry | Yarn |
| 13 | Garments and Other Fiber Products | None assigned |
| 14 | Leather, Furs, Down and Related Products | None assigned |
| 15 | Timber Processing, Bamboo, Cane, Palm & Straw Products | None assigned |
| 16 | Furniture Manufacturing | None assigned |
| 17 | Papermaking and Paper Products | Machine-made paper and paperboard |
| 18 | Printing and Record Medium Reproduction | None assigned |
| 19 | Cultural, Educational and Sports Articles | None assigned |
| 20 | Petroleum Processing and Coking | Crude oil processed |
| 21 | Raw Chemical Materials and Chemical Products | Caustic soda (100% NaOH equivalent) |
| 22 | Medical and Pharmaceutical Products | None assigned |
| 23 | Chemical Fiber | Chemical fibres |
| 24 | Rubber Products | None assigned |
| 25 | Plastic Products | Plastic products |
| 26 | Nonmetal Mineral Products | Cement |
| 27 | Smelting and Pressing of Ferrous Metals | Crude steel |
| 28 | Smelting and Pressing of Nonferrous Metals | Ten non-ferrous metals |
| 29 | Metal Products | None assigned |
| 30 | Ordinary Machinery | Metal-cutting machine tools |
| 31 | Equipment for Special Purposes | Metal smelting equipment |
| 32 | Transportation Equipment | Motor vehicles |
| 33 | Electric Equipment and Machinery | AC motors |
| 34 | Electronic and Telecommunications Equipment | Mobile communication handsets |
| 35 | Instruments, Meters, Cultural and Office Machinery | Electrical measuring instruments |
| 36 | Other Manufacturing Industry | None assigned |
| 37 | Scrap and waste | None assigned |
| 38 | Production and Supply of Gas | None assigned |
| 39 | Production and Supply of Tap Water | None assigned |
| 40 | Construction Industry | None assigned |

**Table S4.** Mapping of the 40 retained industrial and construction subsectors to the five reporting categories, based on Table 2 of Shan et al. [3]

| **Sector** | **Category** |
|---|---|
| Coal Mining and Dressing | Energy production |
| Petroleum and Natural Gas Extraction | Energy production |
| Ferrous Metals Mining and Dressing | Energy production |
| Nonferrous Metals Mining and Dressing | Energy production |
| Non-metal Minerals Mining and Dressing | Energy production |
| Other Minerals Mining and Dressing | Energy production |
| Logging and Transport of Wood and Bamboo | Light manufacturing |
| Food Processing | Light manufacturing |
| Food Production | Light manufacturing |
| Beverage Production | Light manufacturing |
| Tobacco Processing | Light manufacturing |
| Textile Industry | Light manufacturing |
| Garments and Other Fibre Products | Light manufacturing |
| Leather, Furs, Down and Related Products | Light manufacturing |
| Timber Processing, Bamboo, Cane, Palm Fibre & Straw Products | Light manufacturing |
| Furniture Manufacturing | Light manufacturing |
| Papermaking and Paper Products | Light manufacturing |
| Printing and Record Medium Reproduction | Light manufacturing |
| Cultural, Educational and Sports Articles | Light manufacturing |
| Petroleum Processing and Coking | Energy production |
| Raw Chemical Materials and Chemical Products | Heavy manufacturing |
| Medical and Pharmaceutical Products | Light manufacturing |
| Chemical Fibre | Heavy manufacturing |
| Rubber Products | Heavy manufacturing |
| Plastic Products | Heavy manufacturing |
| Non-metal Mineral Products | Heavy manufacturing |
| Smelting and Pressing of Ferrous Metals | Heavy manufacturing |
| Smelting and Pressing of Nonferrous Metals | Heavy manufacturing |
| Metal Products | Heavy manufacturing |
| Ordinary Machinery | Heavy manufacturing |
| Equipment for Special Purposes | Heavy manufacturing |
| Transportation Equipment manufacturing | Heavy manufacturing |
| Electric Equipment and Machinery | High-tech industry |
| Electronic and Telecommunications Equipment | High-tech industry |
| Instruments, Meters, Cultural and Office Machinery | High-tech industry |
| Other Manufacturing Industry | High-tech industry |
| Scrap and waste | High-tech industry |
| Production and Supply of Gas | Energy production |
| Production and Supply of Tap Water | Heavy manufacturing |
| Construction | Construction |

**Table S5.** Sectoral and total provincial uncertainty estimates reproduced from **SI Table 3** supplied with Huang et al. [1]

<table>
<tr><th>Province</th><th>Power</th><th>Industry, Ground Transport, Aviation, Residential</th><th>Total</th></tr>
<tr><td>Hainan</td><td>±23.8%</td><td rowspan="31">±23.8%</td><td>±12.9%</td></tr>
<tr><td>Jilin</td><td>±23.8%</td><td>±13.0%</td></tr>
<tr><td>Shanghai</td><td>±23.8%</td><td>±13.4%</td></tr>
<tr><td>Heilongjiang</td><td>±23.8%</td><td>±13.9%</td></tr>
<tr><td>Chongqing</td><td>±23.8%</td><td>±14.0%</td></tr>
<tr><td>Hunan</td><td>±23.8%</td><td>±14.6%</td></tr>
<tr><td>Fujian</td><td>±23.8%</td><td>±14.8%</td></tr>
<tr><td>Hubei</td><td>±23.8%</td><td>±14.8%</td></tr>
<tr><td>Liaoning</td><td>±23.8%</td><td>±14.8%</td></tr>
<tr><td>Henan</td><td>±23.8%</td><td>±14.9%</td></tr>
<tr><td>Sichuan</td><td>±23.8%</td><td>±14.9%</td></tr>
<tr><td>Jiangxi</td><td>±23.8%</td><td>±15.0%</td></tr>
<tr><td>Zhejiang</td><td>±23.8%</td><td>±15.2%</td></tr>
<tr><td>Jiangsu</td><td>±23.8%</td><td>±15.3%</td></tr>
<tr><td>Guangdong</td><td>±23.8%</td><td>±15.4%</td></tr>
<tr><td>Guangxi</td><td>±23.8%</td><td>±15.5%</td></tr>
<tr><td>Anhui</td><td>±23.8%</td><td>±15.6%</td></tr>
<tr><td>Shandong</td><td>±23.8%</td><td>±15.6%</td></tr>
<tr><td>Guizhou</td><td>±23.8%</td><td>±15.7%</td></tr>
<tr><td>Shanxi</td><td>±23.8%</td><td>±15.9%</td></tr>
<tr><td>Shaanxi</td><td>±23.8%</td><td>±16.2%</td></tr>
<tr><td>Hebei</td><td>±23.8%</td><td>±16.7%</td></tr>
<tr><td>Xizang</td><td>±73.3%</td><td>±16.7%</td></tr>
<tr><td>Ningxia</td><td>±23.8%</td><td>±18.0%</td></tr>
<tr><td>Xinjiang</td><td>±23.8%</td><td>±18.4%</td></tr>
<tr><td>Inner Mongolia</td><td>±23.8%</td><td>±18.8%</td></tr>
<tr><td>Qinghai</td><td>±73.3%</td><td>±20.8%</td></tr>
<tr><td>Yunnan</td><td>±73.3%</td><td>±21.6%</td></tr>
<tr><td>Beijing</td><td>±73.3%</td><td>±29.5%</td></tr>
<tr><td>Tianjin</td><td>±73.3%</td><td>±30.4%</td></tr>
<tr><td>Gansu</td><td>±73.3%</td><td>±38.6%</td></tr>
</table>

**Supplementary Note 1.** Quality control for industrial production indicators and allocation checks.

Monthly industrial production data were processed to obtain consistent monthly activity profiles. Missing monthly values were derived from cumulative production where available. When only January–February cumulative production was reported, the total was divided equally between the two months unless one month's output was available separately. Product series were retained only when all 12 monthly observations were available, values were non-negative, reporting units were consistent, and annual production was positive.

Nine product series with anomalous production concentrated in one or two months were additionally excluded following case-specific review. For subsectors without valid representative product series, monthly emission allocation followed the normalized provincial Carbon Monitor industrial emission profile.

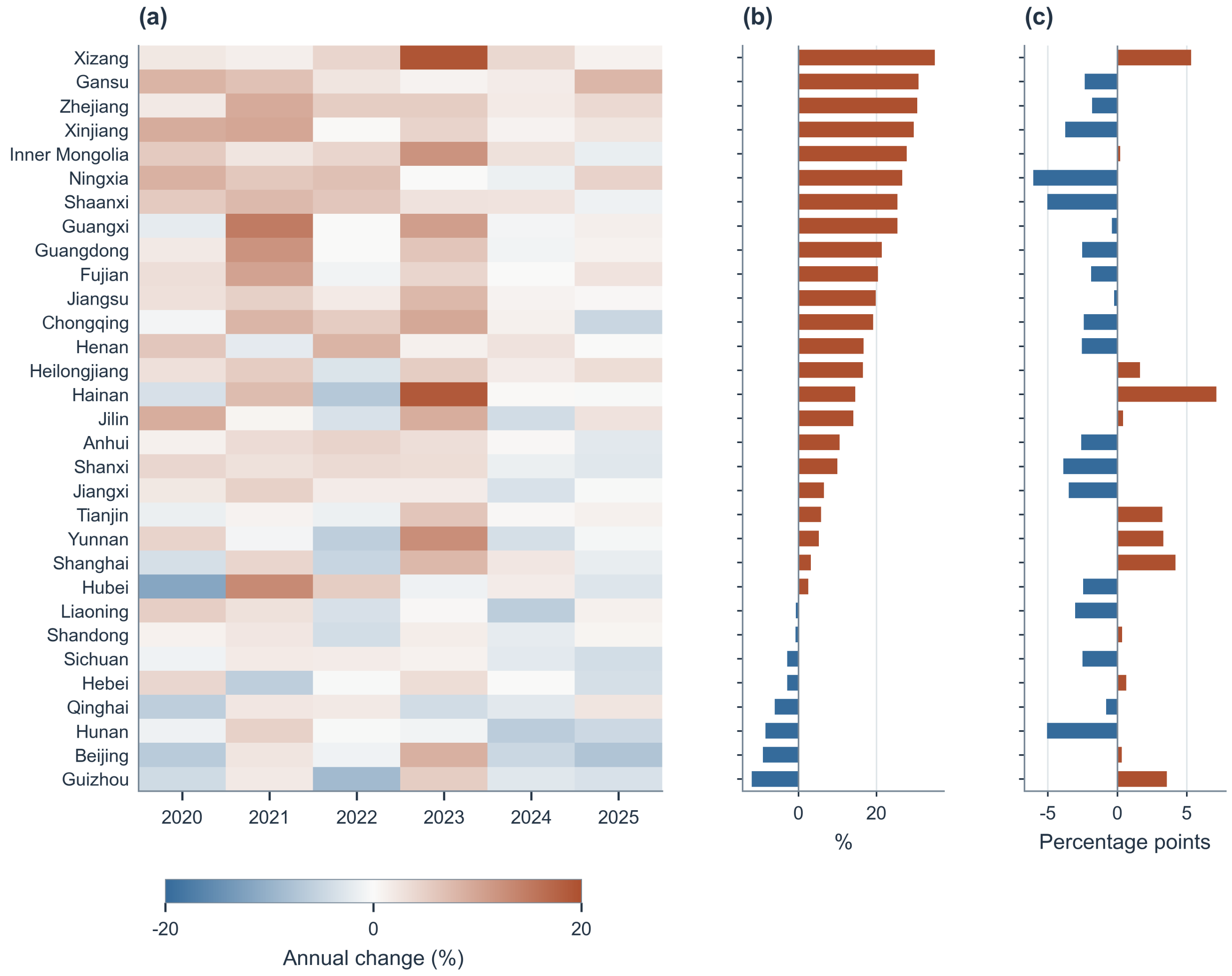

**Figure S1.** Provincial emission trajectories across 31 provincial-level units. (a) Year-on-year percentage change for the six transitions ending in 2020–2025. (b) Percentage change from 2019 to 2025. (c) The difference between the compound annual growth rates for 2022–2025 and 2019–2022. Provinces are ordered by endpoint percentage change.